\documentstyle[preprint,prl,aps]{revtex}

\newcommand{\spc}{{\ }}

\tighten

\begin{document}
\draft
\preprint{STPHY-TH/94-7}

\title   {
             Boson-fermion Dyson mapping and supersymmetry in fermion systems
         }
\author  {  P. Navr\'atil$^{1,2}$, H.B. Geyer$^1$ and
            J. Dobaczewski$^{1,3}$
         }
\address { $^1$Institute of Theoretical Physics, University of Stellenbosch,
                       7599 Stellenbosch, South Africa\\
         $^2$ Institute of Nuclear Physics, Czech Academy of Sciences,
                250 68 \v{R}e\v{z} near Prague, Czech Republic\\
            $^3$Institute of Theoretical Physics, Warsaw University,
                       ul. Ho\.za 69, PL--00-681 Warsaw, Poland
        }

\maketitle

\begin{abstract}

We demonstrate that exact supersymmetry can emerge in a purely
fermionic system.  This ``supersymmetry without bosons" is unveiled by
constructing a novel boson-fermion Dyson mapping from a fermion space
to a space comprised of collective bosons and ideal fermions.  In a
nuclear structure context the collective spectra of even and odd
nuclei can then be unified in a supersymmetric description with Pauli
correlations still exactly taken into account.

\end{abstract}

\pacs{PACS numbers: 21.60.Fw, 21.60.Ev, 11.30.Pb}

\narrowtext

In its original inception supersymmetry pertains to a system of
fermions and bosons exhibiting an invariance with respect to exchange
between these two classes of particles.  It may therefore be somewhat
surprising to discover that a fermion system on its own can also
exhibit {\em exact supersymmetry}.  This is discussed in general below
and demonstrated for a specific nuclear model.

The notion of supersymmetry has in fact proved to be fruitful in
nuclear structure physics \cite{Iac80}.  Properties of some
neighboring even and odd nuclei can namely be classified and
understood in terms of an assumed supersymmetry within the framework
of the interacting boson-fermion model (IBFM) \cite{IvI91}.  To
appreciate how this may come about and why this phenomenological
supersymmetry does not necessarily imply exact supersymmetry on the
microscopic level, we first briefly recapitulate some basics of
nuclear structure.

The atomic nucleus is composed of fermions whose interactions can be
described quite successfully on a non-relativistic level by a fermion
hamiltonian.  States of even and odd nuclei are in principle
eigenstates of the same hamiltonian obtained for different particle
numbers.  Although there is no fundamental difference between even and
odd nuclei from this point of view, their properties are, however,
quite different.

Even-even nuclei, in which numbers of protons and neutrons are both
even, have $0^+$ ground states, $2^+$ first excited states in most
cases and very often low-lying sequences of collective states.
Even-odd nuclei, with an odd number of either neutrons or protons,
have ground states characterized by the spin and parity of the odd
particle, with the particle-core interaction and core deformation
playing a decisive role.

A unification of spectra of even and odd nuclei into a single
framework is therefore a challenging possibility, with the prospect of
unveiling a basic underlying symmetry.  On the phenomenological level
the IBFM \cite{IvI91} achieves such a unification.  Starting from a
common boson-fermion hamiltonian one finds that in some instances
states of an even nucleus, described by many-boson wave functions, are
linked by supersymmetry to states in a neighboring odd nucleus in
which the odd fermion is treated explicitly.  It is important to
realize, however, that Pauli correlations between the odd particle and
the even core are not fully taken into account in the IBFM.  In this
sense the link between the observed supersymmetry and the underlying
microscopy is tenuous and no microscopic derivation of the IBFM
hypothesis in fact exists so far.

In the present paper we systematically derive a supersymmetric nuclear
hamiltonian starting from a purely fermionic microscopic collective
model, while incorporating all Pauli correlations exactly.  We assume
a hamiltonian which is a quadratic scalar function of collective
fermion pair creation operators $A^j$=${\case{1}{2}} \chi^j_{\mu\nu}
a^{\mu} a^{\nu}$, pair annihilation operators $A_i$, and their
commutators $[A_i,A^j]$.  The notation exploits a summation convention
in which upper (lower) indices denote creation (annihilation)
operators, as in $a^{\mu}$=$a^+_{\mu}$ and $A^j$=$A^+_j$.  We also
assume \cite{DGH91,DSG93} that the collective fermion operators obey
the algebraic closure relations
$   \left[\left[A_i,A^j\right],A_k\right] = c^{jl}_{ik}A_l ,$
i.e., operators $A^j$, $A_i$, and $[A_i,A^j]$, to form
a collective spectrum generating algebra \cite{SGA88}.
These relations guarantee an exact decoupling of the even collective
space
  $ |\Psi_{\text{even}}\rangle=A^iA^j\ldots{}A^k|0\rangle$
from all other even fermion states. Similarly, by adding an
odd fermion, one obtains an exactly decoupled odd collective
space
  $ |\Psi_{\text{odd}}\rangle=a^{\mu}A^iA^j\ldots{}A^k|0\rangle$.

In a supersymmetric description, or any boson-fermion description for
that matter, an even state $|\Psi_{\text{even}}\rangle$ should be
represented by an ideal boson space state $|\Psi_{\text{even}}$), say,
with the odd ideal space states $|\Psi_{\text{odd}})$ containing an
additional fermion $\alpha^{\mu}$.  Following traditional terminology
this is also referred to as an ideal fermion \cite{KM91}.  By
definition it commutes with all boson operators,
$[\alpha^{\mu},B_i]$=$[\alpha^{\mu},B^i]$=0.  The required
representation in the ideal space can be achieved by constructing an
appropriate boson-fermion mapping.  In Ref.{\spc}\cite{DSG93}
supercoherent states have been used to derive the boson-fermion
mapping,
   \FL\begin{mathletters}\label{e128}\begin{eqnarray}
   A^j               &\longleftrightarrow& R^j \equiv
                       {\cal{A}}^j  + B^i\big[{\cal{A}}_i,
                       {\cal{A}}^j\big]
                     - {\case{1}{2}}c^{jl}_{ik}B^i B^k B_l
                                             , \label{e128c} \\
   \big[A_i,\!A^j\big] &\longleftrightarrow&
                            \big[{\cal{A}}_i,{\cal{A}}^j\big]
                    - c^{jl}_{ik}B^kB_l
                                              , \label{e128b} \\
   A_j               &\longleftrightarrow&
                            B_j               , \label{e128a} \\
   a^{\nu}           &\longleftrightarrow&
                            \alpha^{\nu}
                    + B^i\big[{\cal{A}}_i,\alpha^{\nu}\big]
                                              ,  \label{e128e} \\
   a_{\nu}           &\longleftrightarrow&
                            \alpha_{\nu}      ,  \label{e128d}
   \end{eqnarray}\end{mathletters}%
where
${\cal{A}}^j$=${\case{1}{2}}\chi^j_{\mu\nu}\alpha^{\mu}\alpha^{\nu}$
are collective pairs of {\em ideal fer\-mions}.  This mapping exactly
preserves the commutation relations of the collective algebra, as well
as the commutation relations between single-fermion and pair
operators, and also the anticommutation relations between
single-fermion operators.

The mapping of states which results from the above mapping of
operators is, however, not satisfactory \cite{DSG93}.  Even states in
particular are not entirely bosonized and two-fermion collective
states are e.g.{\spc}mapped as
$A^j|0\rangle\longleftrightarrow(gB^j$+${\cal{A}}^j$)$|0)$,
where for
$\chi_i^{\mu\nu}=(\chi^i_{\mu\nu})^{\displaystyle\ast}$,
${\case{1}{2}}\chi_i^{\mu\nu}\chi^j_{\mu\nu}=g\delta^j_i$
defines the normalization factor $g$.  In fact, by using ideal
fermions in addition to bosons one has introduced a large redundancy
in the boson-fermion space.  However, this freedom can be used to map
two-fermion states $A^j|0\rangle$ on an arbitrary linear combination
of the one-boson states $B^j|0)$ and one-ideal-fermion-pair states
${\cal{A}}^j|0)$.
A mapping free of the above redundancy can thus be found and
constructed by applying to mapping (\ref{e128}) a suitable similarity
transformation.

Some examples of appropriate similarity transformations have been
presented in Ref.{\spc}\cite{DSG93}.  They led, however, to infinite
series of operators and were rather impractical in applications.  Here
we derive a similarity transformation which does not have this
disadvantage.  The key point of the derivation is the observation that
by simply disregarding the ideal-fermion pair ${\cal{A}}^j$ in the
mapping of the collective pair $A^j$, Eq.{\spc}(\ref{e128c}), we still
preserve the collective commutation relations.  Therefore, there
exists a similarity transformation $X$ of the boson-fermion images
(\ref{e128}) which gives the following mapping:
   \FL\begin{mathletters}\label{e228}\begin{eqnarray}
   A^j               &\longleftrightarrow& R^j - {\cal{A}}^j =
                       B^i\big[{\cal{A}}_i,{\cal{A}}^j\big]
                     - {\case{1}{2}}c^{jl}_{ik}B^i B^k B_l
                                          , \label{e228c} \\
   \big[A_i,\!A^j\big] &\longleftrightarrow&
                            \big[{\cal{A}}_i,{\cal{A}}^j\big]
                    - c^{jl}_{ik}B^kB_l
                                              , \label{e228b} \\
   A_j               &\longleftrightarrow&
                            B_j               , \label{e228a} \\
   a^{\nu}           &\longleftrightarrow&
                            X^{-1}\left(\alpha^{\nu}
                    + B^i\big[{\cal{A}}_i,\alpha^{\nu}\big]\right)X
                                              ,  \label{e228e} \\
   a_{\nu}           &\longleftrightarrow&
                            X^{-1}\alpha_{\nu}X      .  \label{e228d}
   \end{eqnarray}\end{mathletters}%

Before discussing properties of this mapping and its relation to
supersymmetry, we show that the similarity transformation $X$
has the explicit form
\begin{equation}\label{eq3}
    X = \sum_{n=0}^{\infty} \bigl(\frac{1}{C_F-\widehat{C}_F}
        {\cal A}^i B_i\bigr)^n
        {\textstyle \raisebox{-2ex}{$\widehat{}$}} \quad ,
\end{equation}
where $C_F$=${\cal{A}}^k{\cal{A}}_k$ is a Casimir operator of the
ideal-fermion core subalgebra composed of operators
$[{\cal{A}}_i,{\cal{A}}^j]$, i.e.,
$\left[C_F,[{\cal{A}}_i,{\cal{A}}^j]\right]$=0.  Here we use the
notation \cite{Gey86,KV87} of a deferred-action operator
$\widehat{C}_F$ which should be evaluated at the position indicated by
``$\:$\raisebox{-1ex}{$\widehat{}$}$\:$''.  Equivalently, one can
write Eq.{\spc}(\ref{eq3}) by using multiple sums over eigenstates of
$C_F$, in which case $1/(C_F-\widehat{C}_F)$ can be replaced by
typical energy denominators.

First, $X^{-1}B_jX$=$B_j$, because $X$ explicitly commutes with the
boson annihilation operator $B_j$, accounting for (\ref{e228a}).
Second, we show that $X^{-1}R^jX$=$R^j$$-$${\cal{A}}^j$ by proving the
identity
 \begin{equation}\label{e4}
   \left[X,R^j-{\cal{A}}^j\right]={\cal{A}}^jX .
 \end{equation}
This is achieved by splitting $X$ into an infinite sum of terms
which individually increase the number of ideal fermions by 0, 2,
4,$\ldots$,
i.e., $X$=$\sum_{n=0}^{\infty}X_n$. As $R^j$$-$${\cal{A}}^j$
must conserve the number of ideal fermions, Eq.{\spc}(\ref{e4})
becomes the recurrence relation
$   \left[X_{n+1},R^j-{\cal{A}}^j\right]={\cal{A}}^jX_n $
which allows all $X_n$ to be determined, if $X_0$ is known.

We are free to choose $X_0$ as an arbitrary operator which commutes
with $R^j$$-$${\cal{A}}^j$, and Eq.{\spc}(\ref{eq3}) reflects the
simplest choice $X_0$=1.  Then the term $X_1$ in Eq.{\spc}(\ref{eq3})
is a solution of the recurrence relation for $n$=0.  Indeed we have
\widetext
   \begin{equation}\label{e6}
       \bigg[\frac{1}{C_F-\widehat{C}_F} {\cal A}^i B_i\:\:
         {\textstyle \raisebox{-2ex}{$\widehat{}$}}\:\:,
         R^j-{\cal{A}}^j\bigg]
   = \frac{1}{C_F-\widehat{C}_F} \left[{\cal A}^i B_i,
         R^j-{\cal{A}}^j\right]\:
         {\textstyle \raisebox{-2ex}{$\widehat{}$}}
   = \frac{1}{C_F-\widehat{C}_F} (C_F{\cal{A}}^j-{\cal{A}}^jC_F)\:
         {\textstyle \raisebox{-2ex}{$\widehat{}$}} = {\cal{A}}^j,
   \end{equation}
where the first equality results from the fact that $C_F$
commutes with $R^j$$-$${\cal{A}}^j$ and the second one from the
explicit calculation of the commutator. Similarly, we prove by
induction that the $X_{n+1}$ term of Eq.{\spc}(\ref{eq3}) is a
solution of the recurrence relation provided the same is true for
$X_n$:
   \begin{eqnarray}\label{e7}
       [X_{n+1} , R^j-{\cal{A}}^j]
    &=& \frac{1}{C_F-\widehat{C}_F} \left[{\cal{A}}^i B_iX_n,
         R^j-{\cal{A}}^j\right]\:\:
         {\textstyle \raisebox{-2ex}{$\widehat{}$}}\:\:
     = \frac{1}{C_F-\widehat{C}_F} \left({\cal{A}}^i B_i
        {\cal{A}}^jX_{n-1}
         + (C_F{\cal{A}}^j-{\cal{A}}^jC_F)X_n\right)\:\:
         {\textstyle \raisebox{-2ex}{$\widehat{}$}} \nonumber \\
    &=& \:{\textstyle \raisebox{-1ex}{$\breve{}$}}\:\:{\cal{A}}^j
       \frac{1}{\breve{C}_F-\widehat{C}_F}\bigl(1
      +\frac{\breve{C}_F-C_F}{C_F-\widehat{C}_F}\bigr)
       {\cal{A}}^i B_iX_{n-1}\:\:
         {\textstyle \raisebox{-2ex}{$\widehat{}$}}\:\: =
       {\cal{A}}^jX_n
   \end{eqnarray}
where the operator $\breve{C}_F$ is in turn evaluated at
``$\:$\raisebox{-1ex}{$\breve{}$}$\:$''.
\narrowtext

We can now discuss the structure of images of single fermion
operators, Eqs.{\spc}(\ref{e228e}) and (\ref{e228d}).  An explicit
general evaluation these images is difficult because the operators
$\alpha^{\nu}$ and $\alpha_{\nu}$ do not commute with the ideal
fermion Casimir operator $C_F$ and one has to consider branching rules
of the collective algebra.  Before presenting solutions in particular
cases we can, however, analyze some general properties of these
images.

We note that the similarity transformation (\ref{eq3}) does not change
any state in which there are no bosons $B^j$, and therefore the
single-fermion states are mapped onto single ideal fermion states,
$a^{\nu}|0\rangle\leftrightarrow\alpha^{\nu}|0)$.  Since the images of
pair creation operators (\ref{e228c}) do not change the ideal fermion
number, collective odd states $ |\Psi_{\text{odd}}\rangle$ will be
mapped onto ideal states with one ideal fermion only.  Especially from
a physical point of view, this result is a clear improvement over the
solutions found in Ref.{\spc}\cite{DSG93}, where such odd states were
mapped onto ideal states with many-fermion components.

Moreover, the two-fermion states are mapped as
$   a^{\mu}a^{\nu}|0\rangle\longleftrightarrow
   \bigl(\alpha^{\mu}\alpha^{\nu}+\chi^{\mu\nu}_iB^i
     -\frac{1}{C_F}\chi^{\mu\nu}_i{\cal{A}}^i\bigr)|0) $
and in general contain the non-collective pair of ideal fermions
$\alpha^{\mu}\alpha^{\nu}|0)$. However, when the collective pair
$A^j$ is formed by summing the pairs $a^{\mu}a^{\nu}$
with collective amplitudes $\chi^j_{\mu\nu}$, the ideal non-collective
pairs above recombine (note that
$C_F{\cal{A}}^j|0)$=$g{\cal{A}}^j|0)$) and only the boson state
$gB^j|0)$ remains.  Again, since the images (\ref{e228c}) conserve the
ideal fermion number the same recombination mechanism is also valid
for any even state.

When the Casimir operator $C_F$ of a spectrum generating
algebra depends on the ideal fermion number operator only, one may
derive an explicit form of the similarity transformation \cite{NGD94}.
Here we only present the solution for the complete so(2$N$)
algebra of all pairs $a^{\mu}a^{\nu}$ for $\mu$$<$$\nu$,
   \begin{equation}\label{eqso2}
    X = \frac{(n+\hat{n}+1)!!}{(2\hat{n}+1)!!} {\exp}\left(
        {\case{1}{2}}
         \alpha^\mu \alpha^\nu B_{\mu\nu}\right) \:
{\textstyle \raisebox{-2ex}{$\widehat{}$}}\quad,
   \end{equation}
where $n$=$\alpha^{\mu}\alpha_{\mu}$.  This resembles the similarity
transformations considered in Ref.{\spc}\cite{DSG93} with the
important modification of number-operator dependent expansion
coefficients.  The boson-fermion mapping now reads
   \begin{eqnarray}\label{eqsso2n}
   a^{\mu} a^{\nu} &\longleftrightarrow&
                       B^{\mu\nu}
                 - B^{\mu\rho} B^{\nu\theta} B_{\rho\theta}
 - B^{\mu\rho}\alpha^{\nu}\alpha_{\rho}
                     + B^{\nu\rho}\alpha^{\mu}\alpha_{\rho} ,
                \nonumber \\
   a^{\mu} a_{\nu} &\longleftrightarrow&
                       B^{\mu\theta} B_{\nu\theta}
                 + \alpha^{\mu}\alpha_{\nu} ,
                \nonumber \\
   a_{\nu} a_{\mu} &\longleftrightarrow&
                       B_{\mu\nu}               ,
                \label{so110}  \\
   a^{\nu}   &\longleftrightarrow&
           \case{n-2}{2n-3}\alpha^\nu
 + B^{\nu\rho}\alpha_{\rho}
         + \case{1}{2n-1}\alpha^\tau
      B^{\nu\rho}B_{\rho\tau}
                \nonumber \\
   &&  -\case{1}{(2n-1)(2n-3)}\alpha^\sigma\alpha^\tau\alpha_\rho
B^{\nu\rho}B_{\sigma\tau} ,
                \nonumber \\
   a_{\nu}      &\longleftrightarrow&
       \alpha_{\nu}
   -\case{1}{(2n-1)(2n-3)}\alpha^\sigma\alpha^\tau \alpha_\nu B_{\sigma\tau}
   +\case{1}{2n-1}\alpha^\tau B_{\nu\tau} .
                \nonumber
   \end{eqnarray}
The mapping of the bifermion operators is identical to the one
derived by D\"onau and Janssen \cite{DJ73}, while the
mapping of the single-fermion operators completes their result
preserving all fermion (anti-)commutation relations.

Clearly the above mapping (or similar mappings for a collective
algebra\cite{NGD94}) will map a 1-plus-2-body hamiltonian onto a
boson-fermion hamiltonian with the boson-fermion interaction
expressible in terms of the supergenerators $B^\dagger \alpha$ and
$\alpha^\dagger B$. In the ideal space one can then expect
boson-fermion symmetry and under favorable circumstances even
dynamical supersymmetry as we show in more detail for the
boson-fermion mapping of an SO(8) model \cite{Gin80}, defined by
collective monopole and quadrupole fermion pairs $S^+$ and $D^+$.

In order to generalize the model to odd systems we include
$2\Omega$=(2$i$+1)(2$k$+1) creation and annihilation operators
$a^\dagger_{jm}$ and $a_{jm}$, where $j$=$|k$$-$$i|,\ldots,k$+$i$
for an integer value of the inactive angular momentum $k$ and
the active angular momentum $i$=3/2 \cite{Gin80}. A
boson-fermion mapping of this algebra can be derived from
supercoherent states similarly to Eqs.{\spc}(\ref{e128}) \cite{NGD94}.
In this case the series defining the similarity transformation
(\ref{eq3}) cannot be explicitly summed up because the
denominator $C_F-{\hat C}_F$, with
$ C_F = \case{1}{2} n (\Omega+6-\case{1}{2} n)-C_{2 {\rm Spin_F(6)}}$,
contains the quadratic Casimir operator of the Spin(6) group
which is not expressible in terms of number operators.

A system of collective fermion pairs is now mapped onto a system of
$s^{\dag}$ and $d^{\dag}$ bosons, and a system of collective fermion
pairs with an odd fermion onto $s^{\dag}$ and $d^{\dag}$ bosons and
one ideal fermion.  This situation is reminiscent of the
phenomenological IBFM and we proceed by showing that in the SO(8)
model supersymmetry is in fact manifest in the mapped systems.

The six bosons above fix the boson sector of the supersymmetric
structure in terms of U(6).  The size of the fermion sector depends on
$k$, and for $k$=0 ($j$=3/2) the four fermion states lead to an
overall IBFM U(6/4) supersymmetry \cite{IK81,BBI81}.  However, in
SO(8) all particles occupy the same $j$=3/2 level while in the IBFM it
is assumed that the bosons occupy the whole valence shell with only
the fermion restricted to $j$=3/2.

As a more realistic situation, consider $k$=2 corresponding to
$j$=1/2, 3/2, 5/2, and 7/2.  In the IBFM, a related supersymmetry with
the same single-particle content is U(6/20), realized in the Au-Pt
isotopes \cite{L84}.  The group reduction chain is ${\rm U_B(6)}
\otimes {\rm U_F(20)} \supset {\rm SO_B(6)} \otimes {\rm SU_F(4)}
\supset {\rm Spin_{B+F}(6)} \supset {\rm Spin_{B+F}(5)} \supset {\rm
Spin_{B+F}(3)}$.  A Hamiltonian chosen as a linear combination of
quadratic Casimir operators in the chain yields the analytic U(6/20)
IBFM energy formula \cite{L84}
\begin{eqnarray}
E &=& A \sigma(\sigma+4) +
    \tilde{A} [\sigma_1 (\sigma_1 +4)+\sigma_2(\sigma_2+2)+\sigma_3^2]
                                  \nonumber \\
    &+& B[\tau_1(\tau_1+3)+\tau_2(\tau_2+1)]
    + C J(J+1) . \label{eqsy7}
\end{eqnarray}

In the SO(8) Ginocchio model all $j$=1/2, 3/2, 5/2, and 7/2 states are
degenerate which leads to unrealistic odd spectra.  One can lift this
degeneracy by adding to the SO(8) algebra multipole operators
corresponding to interchanged active and inactive angular momenta $k$
and $i$.  Suppose we add two such operators $\bar{P}_J$ for $J$=1 and
3, which form the SO(5) algebra and commute with all SO(8) generators.
We may then consider the fermion group reduction chain ${\rm SO(8)}
\otimes {\rm SO(5)} \supset {\rm Spin(6)} \otimes {\rm SO(5)} \supset
{\rm Spin(5)}\otimes {\rm SO(5)} \supset {\rm \widetilde{Spin}(5)}
\supset {\rm \widetilde{Spin}(3)}$.  Here ${\rm \widetilde{Spin}(5)}$
is generated by $G^{(3)} = \sqrt{5} P_3 - 2\sqrt{2} \bar{P}_3$ and
$G^{(1)} = \sqrt{5} P_1 + 2\sqrt{2} \bar{P}_1$, and ${\rm
\widetilde{Spin}(3)}$ by $G^{(1)}$, where $P_J$ are the original SO(8)
multipole operators.

To arrive at an equivalent boson-fermion description we perform
the SO(8) boson-fermion mapping discussed above, whereas the
new SO(5) algebra is simply mapped from the original fermion
space to the ideal fermion space. It can be shown that the
boson-fermion images of generators of ${\rm
\widetilde{Spin}(5)}$ are (up to a normalization factor) just
the generators of the subgroup Spin(5) of U(6/20) in the IBFM
\cite{ND88}.  Consequently, on the boson-fermion level we have
now the following group chain ${\rm U_B(6)} \otimes {\rm
U_F(20)} \supset {\rm U_B(6)} \otimes {\rm U_F(4)} \otimes {\rm
U_F(5)} \supset {\rm SO_B(6)} \otimes {\rm SU_F(4)} \otimes {\rm
SO_F(5)} \supset {\rm Spin_{B+F}(6)} \otimes {\rm SO_F(5)}
\supset {\rm Spin_{B+F}(5)} \otimes {\rm SO_F(5)} \supset {\rm
\widetilde{Spin}_{B+F}(5)} \supset {\rm \widetilde{Spin}_{B+F}(3)}$.
The associated energy expression is then
\begin{eqnarray}
E &=&   A [\sigma_1 (\sigma_1 +4)+\sigma_2(\sigma_2+2)+\sigma_3^2]
                                        \nonumber \\
  & & + B[\tau_1(\tau_1+3)+\tau_2(\tau_2+1)]
                                        \nonumber \\
  & & + \tilde{B}[\tilde{\tau}_1(\tilde{\tau}_1+3)
    +\tilde{\tau}_2(\tilde{\tau}_2+1)]
    + C J(J+1) , \label{eqsy5}
\end{eqnarray}
where $(\tau_1,\tau_2)$ are the ${\rm Spin_{B+F}(5)}$ irreps
with $\tau_2=1/2$ and $(\tilde{\tau}_1,\tilde{\tau}_2)$ are the
irreps of ${\rm \widetilde{Spin}_{B+F}(5)}$.

As a simple application of the supersymmetric ${\rm
SO(8)}\otimes{\rm SO(5)}$ energy formula (\ref{eqsy5}),
we compare even and odd spectra in
Fig.{\spc}\ref{fig1}.
The Hamiltonian parameters $B=35$ keV, ${\tilde B}=12.63$ keV,
and $C=18$ keV are chosen so that the even part coincides with the
one of Ref.{\spc}\protect{\cite{L84}}. In the odd spectrum we
obtain more low-lying $J$=3/2$^+$ states than one gets in the IBFM
(\ref{eqsy7}).

In summary, we have derived a new generalized Dyson boson-fermion
mapping for a collective algebra extended by single fermion operators.
The mapping gives finite non-hermitian boson-fermion images of
collective pairs and single fermion operators expressed in terms of
ideal boson and fermion annihilation and creation operators.  As an
application we have studied a fermionic model, namely ${\rm
SO(8)}\otimes{\rm SO(5)}$, which extends the SO(8) model by a
non-trivial interaction between the collective pairs and decoupled
particles.  In this model we have revealed a supersymmetric structure
analogous to the interacting boson-fermion model, but with full
recognition of the Pauli principle.

Our study thus shows that supersymmetry, which in principle mixes
bosonic and fermionic degrees of freedom, may equally well appear in
purely fermionic models as it does in purely bosonic models, as
recently demonstrated by Brzezinski {\it et al.} \cite{BEM93} and
Plyushchay \cite{Ply94}.

\bigskip

This work was supported by grants from the Foundation for
Research Development of South Africa, the University of
Stellenbosch, and in part by the Polish State Committee for
Scientific Research under Contract No. 20450~91~01 and by the Grant
Agency of the Czech Republic under grant No. 202/93/2472.

\begin{figure}
\vspace{6.4cm}
\caption{
The lower part of an SO(8)$\otimes$SO(5) spectrum.  In the odd part
(right), the (1/2,1/2) and (3/2,1/2) irreps of Spin(5) are shown only.
The dotted lines connect states belonging to the same ${\rm
\widetilde{Spin}(5)}$ irrep.
}
\label{fig1}
\end{figure}

\end{document}